\documentclass{article}
\usepackage{amsmath,amssymb}
\usepackage{enumerate}

\begin{document}
\title{Invariant Variation Problems}
\author{Emmy Noether} 
\date{} 
\maketitle

M. A. Tavel's English translation of ``Invariante
Variationsprobleme,''
\emph{Nachr. d. K\"{o}nig. Gesellsch. d. Wiss. zu G\"{o}ttingen,
Math-phys. Klasse}, 235--257 (1918), which originally appeared in
\emph{Transport Theory and Statistical Physics}, \textbf{1} (3),
183--207 (1971).\footnote[0]{This paper is reproduced by Frank Y. Wang 
(\texttt{fwang@lagcc.cuny.edu}) with \LaTeX.}

\begin{abstract}
The problems in variation here concerned are such as to admit a
continuous group (in Lie's sense); the conclusions that emerge from
the corresponding differential equations find their most general
expression in the theorems formulated in Section 1 and proved in
following sections.  Concerning these differential equations that
arise from problems of variation, far more precise statements can be
made than about arbitrary differential equations admitting of a group,
which are the subject of Lie's researches.  What is to follow,
therefore, represents a combination of the methods of the formal
calculus of variations with those of Lie's group theory.  For special
groups and problems in variation, this combination of methods is not
new; I may cite Hamel and Herglotz for special finite groups, Lorentz
and his pupils (for instance Fokker), Weyl and Klein for special
infinite groups.\footnote{Hamel, Math. Ann. \textbf{59} and
Z. f. Math. u. Phys. \textbf{50}.  Herglotz, Ann. d. Phys. (4)
\textbf{36}, esp. \S~9, p. 511.  Fokker, Verslag d. Amsterdamer
Akad. Jan. 27, 1917.  For further bibliography, compare Klein's second
Note, G\"{o}ttinger Nachrichten, July 19, 1918.  The recently
published work by Kneser (Math. Zschr. 2) deals with the setting up of
invariants by a similar method.} Especially Klein's second Note and
the present developments have been mutually influenced by each other,
in which regard I may refer to the concluding remarks of Klein's Note.
\end{abstract}

\section{Preliminary Remarks and Formulation of Theorems}

All functions occurring in the sequel are to be assumed analytic, or
at least continuous and continuously differentiable a definite number
of times, and unique in the interval considered.  

By a ``group of transformation,'' familiarly, is meant a system of
transformations such that for each transformation, there exists an
inverse contained in the system, and such that the composition of any
two transformations of the system in turn belongs to the system.  The
group will be called a finite continuous group $ \mathfrak{G}_{\rho}$
if its transformations are contained in a most general
(transformation) depending analytically on $\rho$ essential parameters
$\epsilon$ (i.e., the $\rho$ parameters are not to be representable as
$\rho$ function of fewer parameters).  Correspondingly, an infinite
continuous group $\mathfrak{G}_{\infty \rho}$ is understood to be a
group whose most general transformations depend on $\rho$ essential
arbitrary functions $p(x)$ and their derivatives analytically, or at
least in a continuous and finite-fold continuously differentiable
manner.  The group depending on infinitely many parameters but not on
arbitrary functions stands as an intermediate term between the two.
Finally, a group depending both on arbitrary functions and on
parameters is called a mixed group.\footnote{Lie, in ``Grundlagen
f\"{u}r die Theorie der Unendlichen kontinuierlichen
Transformationsgruppen'' (Foundations of the theory of infinite
continuous groups of transformations),
Ber. d. K. Sachs. Ges. d. Wissensch 1981 (cited as Grundlagen),
defines the infinite continuous group as a group of transformations
which are given by the most general solutions of a system of partial
differential equations, provided these solutions do not depend only on
a finite number of parameters.  One of the above-mentioned types
differing from the finite group will thus be thereby obtained; whereas
conversely the limiting case of infinitely many parameters need not
necessarily satisfy a system of differential equations.}

Let $x_{1}, \ldots, x_{n}$ be independent variables and $u_{1}(x),
\ldots, u_{\mu}(x)$ functions depending upon them.  If the $x$'s and
$u$'s are subjected to the transformations of a group, then, by
hypothesis of invertibility of the transformations, there must again
be exactly $n$ independent variables $y_{1}, \ldots, y_{n}$ among the
transformed quantities; let the others depending upon them be
designated by $v_{1}(y), \ldots, v_{\mu}(y)$.  In the transformations,
the derivatives of the $u$'s with respect to the $x$'s, namely
$\displaystyle{\frac{\partial u}{\partial x}, \frac{\partial^{2}
u}{\partial x^{2}}, \ldots}$ may also occur.\footnote{I suppress the
subscripts, insofar as feasible, even in summations; thus,
$\displaystyle{\frac{\partial^{2} u}{\partial x^{2}}}$ for
$\displaystyle{\frac{\partial^{2} u_{\alpha}}{\partial x_{\beta}
\partial x_{\gamma}}}$, etc.}  A function is called an invariant of
the group if there subsists a relationship
$$
P \left(x, u, \frac{\partial u}{\partial x}, 
\frac{\partial^{2} u}{\partial x^{2}}, \ldots
\right)
=
P \left(y, v, \frac{\partial v}{\partial y}, 
\frac{\partial^{2} v}{\partial y^{2}}, \ldots
\right) .
$$
In particular, then, an integral $I$ will be an invariant of the group
if there subsists a relationship
\begin{equation}
I = \int \ldots \int f \left( x, u, \frac{\partial u}{\partial x}, 
\frac{\partial^{2} u}{\partial x^{2}}
\ldots
\right) d x
= \int \ldots \int f \left( y, v, \frac{\partial v}{\partial y}, 
\frac{\partial^{2} v}{\partial y^{2}}
\ldots
\right) d y
\ \ \ \ \
\footnote{By way of abbreviation, I write $dx$, $dy$ for $dx_{1}
\ldots dx_{n}$, $dy_{1} \ldots dy_{n}$.}
\end{equation}
integrated over an \emph{arbitrary} real $x$-interval and the
corresponding $y$-interval.\footnote{All arguments $x$, $u$,
$\epsilon$, $p(x)$ occurring in the transformations are to be assumed
real, whereas the coefficients may be complex.  But since the final
results are concerned with \emph{identities} in the $x$'s, $u$'s,
parameters and arbitrary functions, these hold also for complex
values, provided only that all functions that occur are assumed
analytic.  A large portion of the results, incidentally, can be
justified without integrals, so that here the restriction to reals is
not necessary even to the arguments.  On the other hand, the
developments at the close of Section 2 and beginning of Section 5 do
not appear to be feasible without integrals.}

Then, for an arbitrary, not necessarily invariant integral $I$, I
form the first variation $\delta I$ and transform it by partial
integration according to the rules of the calculus of variations.  As
we know, provided $\delta u$ with all derivatives that occur is
assumed to vanish at the boundary, while otherwise arbitrary,
\begin{equation}
\delta I = \int \ldots \int \delta f \, dx =
\int \ldots \int 
\left( \sum \psi_{i} \left(x, u, \frac{\partial u}{\partial x}, 
\ldots \right) \delta u_{i} \right) dx ,
\end{equation}
where $\psi$ stands for the Lagrange expressions, i.e., the left-hand
sides of the Lagrange equations of the corresponding variation problem
$\delta I=0$.  To this integral relationship there corresponds an
integral-free identity in $\delta u$ and its derivatives, generated by
writing in the boundary terms as well.  As the partial integration
shows, these boundary terms are integrals over divergences, i.e., over
expressions
$$
\operatorname{Div} A = \frac{\partial A_{1}}{\partial x_{1}} + \ldots + 
\frac{\partial A_{n}}{\partial x_{n}} ,
$$
where $A$ is linear in $\delta u$ and its derivatives.  Hence
\begin{equation}
\sum \psi_{i} \, \delta u_{i} = \delta f + \operatorname{Div} A .
\end{equation}
If in particular $f$ contains only first derivatives of the $u$'s,
then in the case of the single integral the identity (3) is identical
with what Heun calls the ``central equation of Lagrange'' 
\begin{equation}
\sum \psi_{i} \, \delta u_{i} = \delta f - \frac{d}{dx} \left(\sum
\frac{\partial f}{\partial u_{i}^{\prime}} \delta u_{i} \right) ,
\quad
\left(u_{i}^{\prime} = \frac{d u_{i}}{d x} \right) ,
\end{equation}
whereas for the $n$-fold integral, (3) goes over into
\begin{equation}
\sum \psi_{i} \, \delta u_{i} = \delta f -
\frac{\partial}{\partial x_{1}} 
\left(\sum \frac{\partial f}
{\partial \frac{\partial u_{i}}{\partial x_{1}}} \delta u_{i} \right)
- \ldots -
\frac{\partial}{\partial x_{n}} 
\left(\sum \frac{\partial f}
{\partial \frac{\partial u_{i}}{\partial x_{n}}} \delta u_{i} \right) .
\end{equation}
For the single integral and $\kappa$ derivatives of the $u$'s, (3) is
given by
\begin{multline}
\sum \psi_{i} \, \delta u_{i} = \delta f - \\
- \frac{d}{dx} \left\{ \sum\left( 
{1 \choose 1}
\frac{\partial f}{\partial u_{i}^{(1)}} \delta u_{i} +
{2 \choose 1}
\frac{\partial f}{\partial u_{i}^{(2)}} \delta u_{i}^{(1)} +
\ldots
+ {\kappa \choose 1} 
\frac{\partial f}{\partial u_{i}^{(\kappa)}} \delta u_{i}^{(\kappa-1)} 
\right) \right\} +
\\
+ \frac{d^{2}}{dx^{2}} \left\{ \sum\left( 
{2 \choose 2}
\frac{\partial f}{\partial u_{i}^{(2)}} \delta u_{i} +
{3 \choose 2}
\frac{\partial f}{\partial u_{i}^{(3)}} \delta u_{i}^{(1)} +
\ldots
+ {\kappa \choose 2} 
\frac{\partial f}{\partial u_{i}^{(\kappa)}} \delta u_{i}^{(\kappa-2)} 
\right) \right\} + \ldots
\\
+
(-1)^{\kappa} \frac{d^{\kappa}}{d x^{\kappa}} \left\{ \sum 
{\kappa \choose \kappa} 
\frac{\partial f}{\partial u_{i}^{(\kappa)}} \delta u_{i}
\right\}
\end{multline}
and a corresponding identity holds for the $n$-fold integral; in
particular, $A$ contains $\delta u$ as far as the $(\kappa
-1)$st derivative.  The fact that (4), (5), and (6) actually define
the Lagrange expressions $\psi_{i}$ follows from the fact that the
combinations of the right-hand sides eliminate all higher derivatives
of the $\delta u$'s, while on the other hand the relation (2), to
which the partial integration leads \emph{uniquely}, is satisfied.

Now in the following we shall be concerned with these two theorems:
\begin{enumerate}[I.]
\item
If the integral $I$ is invariant with respect to a
$\mathfrak{G}_{\rho}$, then $\rho$ linearly independent combinations
of the Lagrange expressions become divergences --- and from this,
conversely, invariance of $I$ with respect to a $\mathfrak{G}_{\rho}$
will follow.  The theorem holds good even in the limiting case of
infinitely many parameters.
\item
If the integral $I$ is invariant with respect to a
$\mathfrak{G}_{\infty \rho}$ in which the arbitrary functions occur up
to the $\sigma$-th derivative, then there subsist $\rho$ identity
relationships between the Lagrange expressions and their derivatives
up to the $\sigma$-th order.  In this case also, the converse
holds.\footnote{For certain trivial exceptions, compare Section 2,
Note 13.}
%Section 2, 2nd remark.}
\end{enumerate}

For mixed groups, the statements of both theorems hold; that is, both
dependencies and divergence relations independent thereof occur.

Passing over from these identities to the corresponding variation
problem, i.e., putting $\psi=0$,\footnote{Somewhat more generally, we
may alternatively put $\psi_{i}=T_{i}$; cf. Section 3, Note 15.}
Theorem I in the one-dimensional case --- where the divergence goes
over into a total differential --- asserts the existence of $\rho$
first integrals, between which, however, non-linear dependencies may
subsist;\footnote{Cf. close of Section 3.}  in the multidimensional
case, the divergence equations often referred to of late as ``laws of
conservation'' are obtained; Theorem II states that $\rho$ of the
Lagrange equations are a consequence of the rest.

The simplest example of Theorem II --- without converse --- is
afforded by the Weierstrass parametric representation; here the
integral, with homogeneity of first order, is as we know invariant if
the independent variable $x$ is replaced by an arbitrary function of
$x$ that leaves $u$ unchanged ($y=p(x)$; $v_{i}(y)=u_{i}(x)$).  Thus
one arbitrary function occurs, but without derivatives, and to this
corresponds the known linear relationship among the Lagrange
expressions themselves $\displaystyle{\sum \psi_{i} \frac{d u_{i}}{dx}
= 0}$.  Another example is presented by the ``general theory of
relativity'' of the physicists; there we have the group of \emph{all}
transformations $y_{i}=p_{i}(x)$ of the $x$'s, while the $u$'s
(designated as $g_{\mu \nu}$ and $q$) are subjected to the
transformations thereby induced for the coefficients of a quadratic
and linear differential form --- transformations which contain the
first derivatives of the arbitrary function $p(x)$.  To this
correspond the familiar $n$ dependencies between the Lagrange
expressions and their first derivatives.\footnote{Cf. e.g., Klein's
presentation.}

If in particular we specialize the group by allowing no derivatives of
the $u(x)$'s in the transformations, and moreover let the transformed
independent quantities depend only on the $x$'s, not on the $u$'s,
then (as is shown in Section 5) the invariance of $I$ entails the
relative invariance of $\sum \psi_{i} \delta u_{i}$,\footnote{That is,
$\sum \psi_{i} \delta u_{i}$ acquires a factor upon transformation.}
and likewise of the divergences occurring in Theorem I, once the
parameters are subjected to suitable transformations.  For Theorem II,
similarly, we get relative invariance of the left-hand sides of the
dependencies as associated with the aid of the arbitrary functions;
and as a consequence of this, another function whose divergence
vanishes identically and admits of the group --- mediating, in the
physicists' theory of relativity, the connection between dependencies
and the law of conservation of energy.\footnote{Compare Klein's second
note.} Theorem II, finally, in terms of group theory, furnishes the
proof of a related Hilbertian assertion about the failure of laws of
conservation of energy proper in ``general relativity.''  With these
supplementary remarks, Theorem I comprises all theorems on first
integrals known to mechanics etc., while Theorem II may be described
as the utmost possible generalization of the ``general theory of
relativity'' in group theory.

\section{Divergence Relationships and Dependencies}

Let $\mathfrak{G}$ be a --- finite or infinite --- continuous group;
then it is always possible to arrange for the zero values of the
parameters $\epsilon$, or of the arbitrary function $p(x)$, to
correspond to the identity transformation.\footnote{Cf. e.g., Lie,
Grundlagen, p. 331.  Where arbitrary functions are concerned, the
special values $a^{\sigma}$ of the parameters are to be replaced by
fixed functions $p^{\sigma}$, $\displaystyle{\frac{\partial
p^{\sigma}}{\partial x}}$, $\ldots$; and correspondingly, the values
$a^{\sigma}+\epsilon$ by $p + p(x)$, $\displaystyle{\frac{\partial
p^{\sigma}}{\partial x} + \frac{\partial p}{\partial x}}$, etc.}  The
most general transformation will therefore be of the form
$$
y_{i} = A_{i} \left(x, u, \frac{\partial u}{\partial x}, \ldots \right) =
x_{i} + \Delta x_{i} + \ldots
$$
$$
v_{i}(y) = B_{i} \left(x, u, \frac{\partial u}{\partial x}, \ldots \right) =
u_{i} + \Delta u_{i} + \ldots 
$$
where $\Delta x_{i}$, $\Delta u_{i}$ stand for the terms of lowest
dimension in $\epsilon$, or $p(x)$ and its derivatives; in which, in
fact, they will be assumed linear.  As will afterwards appear, this is
no restriction of generality.

Now let the integral $I$ be an invariant with respect to
$\mathfrak{G}$, satisfying, that is, the relationship (1).  Then in
particular, $I$ will also be invariant with respect to the
infinitesimal transformation
$$
y_{i} = x_{i} + \Delta x_{i} ; \quad
v_{i}(y) = u_{i} + \Delta u_{i} ;
$$
contained in $\mathfrak{G}$, and for this relation (1) goes over into
\begin{equation}
0 = \Delta I = 
\int \ldots \int f \left(y, v(y), \frac{\partial
v}{\partial y}, \ldots \right) dy -
\int \ldots \int f \left(x, u(x), \frac{\partial
u}{\partial x}, \ldots \right) dx ,
\end{equation}
where the first integral is to be extended over the $x+\Delta x$
interval corresponding to the $x$-interval.  But this integration may
alternatively be transformed into an integration over the
$x$-interval, by virtue of the transformation, valid for
infinitesimal $\Delta x$,
\begin{equation}
\int \ldots \int f \left(y, v(y), \frac{\partial
v}{\partial y}, \ldots \right) dy =
\int \ldots \int f \left(x, v(x), \frac{\partial
v}{\partial x}, \ldots \right) dx +
\int \ldots \int \operatorname{Div} \left(f \cdot \Delta x \right) d x.
\end{equation}
So if in place of the infinitesimal transformation $\Delta u$, we
introduce the variation
\begin{equation}
\overline{\delta} u_{i} = v_{i}(x) - u_{i}(x)
= \Delta u_{i} - \sum 
\frac{\partial u_{i}}{\partial x_{\lambda}} \Delta x_{\lambda} ,
\end{equation}
then (7) and (8) go over into
\begin{equation}
0 = \int \ldots \int \left\{ \overline{\delta} f + 
\operatorname{Div} (f \cdot \Delta x) \right\} d x .
\end{equation}
The right-hand side is the familiar formula for simultaneous variation
of the dependent and independent variables.  Since the relation (10)
is satisfied for integration over any arbitrary interval, the
integrand must vanish identically; therefore Lie's differential
equations for the invariance of $I$ goes over into the relation
\begin{equation}
\overline{\delta} f + \operatorname{Div} ( f \cdot \Delta x) = 0 .
\end{equation}
If in this, by (3), $\overline{\delta}f$ is expressed in terms of the
Lagrange expressions, we get
\begin{equation}
\sum \psi_{i} \overline{\delta} u_{i} = \operatorname{Div} B \quad
(B = A - f \cdot \Delta x ) ,
\end{equation}
and this relationship, therefore, for every invariant integral $I$,
represents an identity in all arguments that occur; it is the required
form of Lie's differential equations for $I$.\footnote{(12) goes over
into $0=0$ for the trivial case --- which can occur only if $\Delta
x$, $\Delta u$ depend also on derivatives of the $u$'s --- when
$\operatorname{Div} (f \cdot \Delta x)=0$, $\overline{\delta} u = 0$;
thus these infinitesimal transformations are always to be eliminated
from the groups, and only the number of remaining parameters, or
arbitrary functions, is to be counted in the formulation of the
theorems.  Whether the remaining infinitesimal transformations still
form a group must be left moot.}

Now for the present let $\mathfrak{G}$ be taken to be a finite
continuous group $\mathfrak{G}_{\rho}$; since by hypothesis $\Delta u$
and $\Delta x$ are linear in the parameters $\epsilon_{1}$, $\ldots$,
$\epsilon_{\rho}$, hence by (9) the same holds for $\overline{\delta}
u$ and its derivatives; therefore $A$ and $B$ are linear in the
$\epsilon$'s.  So if I let
$$
B = B^{(1)} \epsilon_{1} + \ldots + B^{(\rho)} \epsilon_{\rho} ; \quad
\overline{\delta} u = \overline{\delta} u^{(1)} \epsilon_{1} + \ldots +
\overline{\delta} u^{(\rho)} \epsilon_{\rho} ,
$$
where, that is $\overline{\delta} u^{(1)}$, $\ldots$ are functions of
$x$, $u$, $\displaystyle{\frac{\partial u}{\partial x}}$, $\ldots$,
the required divergence relationships follow from (12):
\begin{equation}
\sum \psi_{i} \overline{\delta} u_{i}^{(1)} = \operatorname{Div} B^{(1)} ; 
\quad \ldots \quad
\sum \psi_{i} \overline{\delta} u_{i}^{(\rho)} = \operatorname{Div}
B^{(\rho)} .
\end{equation}
Thus $\rho$ linearly independent combinations of the Lagrange
expressions become divergences; the linear independence follows from
the fact that by (9), $\overline{\delta} u = 0$, $\Delta x = 0$ would
entail $\Delta u = 0$, $\Delta x = 0$, or in other words a dependency
between the infinitesimal transformations.  But by hypothesis, none
such is satisfied for any value of the parameters, since otherwise the
$\mathfrak{G}_{\rho}$ regenerated by integration from the
infinitesimal transformations would depend on fewer than $\rho$
essential parameters.  But the further possibility $\overline{\delta}
u = 0$, $\operatorname{Div} (f \cdot \Delta x) = 0$ was excluded.
These conclusions hold good even in the limiting case of infinitely
many parameters.

Now let $\mathfrak{G}$ be an infinite continuous group
$\mathfrak{G}_{\infty \rho}$; then $\overline{\delta} u$ and its
derivatives, and hence $B$ also, will again be linear in the arbitrary
functions of $p(x)$ and their derivatives;\footnote{That it signifies
no restriction to assume the $p$'s free from $u$,
$\displaystyle{\frac{\partial u}{\partial x}}$, is shown by the
converse.} independently of (12), further, by substitution of the
values of $\overline{\delta} u$, let
$$
\sum \psi_{i} \overline{\delta} u_{i} = 
\sum_{\lambda, i} \psi_{i} 
\left\{a_{i}^{(\lambda)}(x, u, \ldots) p^{(\lambda)}(x) +
b_{i}^{(\lambda)}(x, u, \ldots) 
\frac{\partial p^{(\lambda)}}{\partial x} + \ldots +
c_{i}^{(\lambda)}(x, u, \ldots) 
\frac{\partial^{\sigma} p^{(\lambda)}}{\partial x^{\sigma}}  
\right\} .
$$
Now, by the identity
$$
\varphi(x, u, \ldots) \frac{\partial^{\tau} p(x)}{\partial x^{\tau}}
=(-1)^{\tau} \cdot
\frac{\partial^{\tau} \varphi}{\partial x^{\tau}} 
\cdot p(x) \ 
\bmod \operatorname{Divergences}
$$
and analogously to the partial integration formula, the derivatives of
$p$ can be replaced by $p$ itself and by divergences that will be
linear in $p$ and its derivatives; hence we get
\begin{equation}
\sum \psi_{i} \overline{\delta} u_{i} = 
\sum_{\lambda} \left\{ (a_{i}^{(\lambda)} \psi_{i}) -
\frac{\partial}{\partial x} (b_{i}^{(\lambda)} \psi_{i}) + \ldots +
(-1)^{\sigma} \frac{\partial^{\sigma}}{\partial x^{\sigma}}
(c_{i}^{(\lambda)} \psi_{i}) 
\right\} p^{(\lambda)} + \operatorname{Div} \Gamma
\end{equation}
and in conjunction with (12)
\begin{equation}
\sum \left\{ (a_{i}^{(\lambda)} \psi_{i}) -
\frac{\partial}{\partial x} (b_{i}^{(\lambda)} \psi_{i}) + \ldots +
(-1)^{\sigma} \frac{\partial^{\sigma}}{\partial x^{\sigma}}
(c_{i}^{(\lambda)} \psi_{i}) 
\right\} p^{(\lambda)} = \operatorname{Div} (B-\Gamma) .
\end{equation}
I now form the $n$-fold integral over (15), extended over any
interval; and choose the $p(x)$'s such that they, with all derivatives
occurring in ($B-\Gamma$), will vanish at the boundary.  Since the
integral over a divergence reduces to a boundary integral, then, the
integral over the left side of (15) will also vanish for $p(x)$'s
which are arbitrary except that they and sufficiently many of their
derivatives vanish at the boundary; and thence follows, by known
inferences, the vanishing of the integrand for every $p(x)$, or in
other words the $\rho$ relationships:
\begin{equation}
\sum \left\{ (a_{i}^{(\lambda)} \psi_{i}) -
\frac{\partial}{\partial x} (b_{i}^{(\lambda)} \psi_{i}) + \ldots +
(-1)^{\sigma} \frac{\partial^{\sigma}}{\partial x^{\sigma}}
(c_{i}^{(\lambda)} \psi_{i}) 
\right\}  = 0 \quad (\lambda = 1, 2, \ldots, \rho).
\end{equation}
These are the required dependencies between the Lagrange expressions
and their derivatives for invariance of $I$ with respect to
$\mathfrak{G}_{\infty \rho}$; the linear independence is proved as
above, since the converse leads back to (12), and since we can again
argue back from the infinitesimal transformations to the finite ones,
as will be explained more fully in Section 4.  In the case of a
$\mathfrak{G}_{\infty \rho}$, that is to say, even in the
infinitesimal transformations there always occur $\rho$ arbitrary
transformations.  Equation (15) and (16) further entail
$\operatorname{Div} (B-\Gamma) = 0$.

If, as corresponds to a ``mixed group,'' $\Delta x$ and $\Delta u$ are
taken linear in the $\epsilon$'s and $p(x)$'s, then we see, by
equating first the $p(x)$'s and then the $\epsilon$'s to zero, that
both divergence relationships (13) and dependencies (16) hold.

\section{Converse in Case of Finite Group}

To prove the converse, we are first to run through essentially the
foregoing arguments in reverse order.  From the fact of (13), by
multiplication by the $\epsilon$'s and adding, the fact of (12)
follows; and thence, by virtue of the identity (3), a relationship
$\overline{\delta} f + \operatorname{Div} (A-B) = 0$.  So if we put
$\displaystyle{\Delta x = \frac{1}{f} (A - B)}$, we have thereby
arrived at (11); whence finally, by integration, there follows (7),
$\Delta I = 0$, or in other words the invariance of $I$ with respect
to the infinitesimal transformation determined by $\Delta x$, $\Delta
u$, where $\Delta u$'s by virtue of (9) are determined from $\Delta x$
and $\overline{\delta} u$, and $\Delta x$ and $\Delta u$ become linear
in the parameters.  But $\Delta I = 0$ entails, in known manner, the
invariance of $I$ with respect to the finite transformations generated
by integration of the simultaneous system
\begin{equation}
\frac{dx_{i}}{dt} = \Delta x_{i} ; \quad
\frac{du_{i}}{dt} = \Delta u_{i} ; \quad
(x_{i} = y_{i}, \ u_{i} = v_{i}, \ \text{for} \ t = 0 ) .
\end{equation} 
These finite transformations contain $\rho$ parameters $a_{1} \ldots
a_{\rho}$, namely the combinations $t \epsilon_{1}$, $\ldots$, $t
\epsilon_{\rho}$.  From the assumption that there are $\rho$ and only
$\rho$ linearly independent divergence relationships (13), it follows
further that the finite transformations, once they do not contain the
derivatives $\displaystyle{\frac{\partial u}{\partial x}}$, always
form a group.  For in the contrary case, at least one infinitesimal
transformation generated by Lie's bracketing process would fail to be
a linear combination of the other $\rho$; and since $I$ admits of this
transformation also, there would be more than $\rho$ linearly
independent divergence relationships; or else that infinitesimal
transformation would be of the special form where $\overline{\delta} u
= 0$, $\operatorname{Div} (f \cdot \Delta x) = 0$, but in that case
$\Delta x$ or $\Delta u$, contrary to hypothesis, would depend on
derivatives.  Whether this case can arise when derivatives occur in
$\Delta x$ or $\Delta u$ must be left moot; in that case, the $\Delta
x$ determined above must be augmented by all functions $\Delta x$ for
which $\operatorname{Div} (f \cdot \Delta x) = 0$ to restore the
group property, but by agreement the parameters thereby adjoined are
not to count.  This completes the proof of the converse.

From this conversion, it follows further that $\Delta x$ and $\Delta
u$ can actually be assumed linear in the parameters.  For if $\Delta
u$ and $\Delta x$ were forms of higher degree in $\epsilon$, then by
the linear independence of the power products of the $\epsilon$'s,
quite analogous relations (13) would follow, only in greater number,
from which, by the converse, invariance of $I$ follows with respect to
a group whose infinitesimal transformations contain the parameters
\emph{linearly}.  If this group is to contain exactly $\rho$
parameters, then linear dependencies must subsist between the
divergence relationships originally obtained through the terms of
higher degree in $\epsilon$.

Let us add the remark that in the case where $\Delta x$ and $\Delta u$
also contain derivatives of the $u$'s, the finite transformations may
depend on infinitely many derivatives of the $u$'s; for in that case
the integration of (17), in the determination of
$\displaystyle{\frac{d^{2} x_{i}}{d t^{2}}}$,
$\displaystyle{\frac{d^{2} u_{i}}{d t^{2}}}$ leads to
$\displaystyle{\Delta \left(\frac{\partial u}{\partial
x_{\kappa}}\right) = \frac{\partial \Delta u}{\partial x_{\kappa}} -
\sum_{\lambda} \frac{\partial u}{\partial x_{\lambda}} \frac{\partial
\Delta x_{\lambda}}{\partial x_{\kappa}}}$, so that the number of
derivatives in general increases at each step.  By way of example,
say,
$$
f = \frac{1}{2} u^{\prime 2} ; \quad
\psi = - u^{\prime \prime} ; \quad
\psi \cdot x = \frac{d}{dx} (u - u^{\prime} x) ; \quad
\overline{\delta} u = x \cdot \epsilon;
$$
$$
\Delta x = \frac{-2 u}{u^{\prime 2}} \epsilon ; \quad
\Delta u = \left(x-\frac{2 u}{u^{\prime}} \right) \cdot \epsilon 
$$
Since the Lagrange expressions of a divergence vanish identically, the
converse shows, finally, the following: if $I$ admits of a
$\mathfrak{G}_{\rho}$, then any integral that differs from $I$ only by
a boundary integral, i.e., by an integral over a divergence, likewise
admits of a $\mathfrak{G}_{\rho}$ having the same $\overline{\delta}
u$'s whose infinitesimal transformations will in general contain
derivatives of the $u$'s.  Thus for instance, corresponding to the
above example, 
%(Tavel p 195) 
$\displaystyle{f^{\ast} = \frac{1}{2} \left\{u^{\prime 2} -
\frac{d}{dx} \left(\frac{u^{2}}{x} \right) \right\}}$ admits of the
infinitesimal transformation $\Delta u = x \epsilon$, $\Delta x = 0$;
while derivatives of the $u$'s occur in the infinitesimal
transformations corresponding to $f$.

Passing over to the variations problem, i.e., putting
$\psi_{i}=0$,\footnote{$\psi_{i}=0$, or, somewhat more generally,
$\psi_{i}=T_{i}$, where $T_{i}$ are newly adjoined functions, are
referred to in physics as ``field equations.''  In the case
$\psi_{i}=T_{i}$, the identities (13) goes over into equations
$\operatorname{Div} B^{(\lambda)} = \sum T_{i} \delta
u_{i}^{(\lambda)}$, likewise known in physics as laws of
conservation.} (13) goes over into the equation $\operatorname{Div}
B^{(1)} = 0, \ldots, \operatorname{Div} B^{(\rho)} = 0$, often
referred to as ``laws of conservation.''  In the one-dimensional case,
it follows from this that $B^{(1)} = \operatorname{const.}$, $B^{(\rho)} =
\operatorname{const.}$; and here the $B$'s contain at most $(2 \kappa - 1)$st
derivatives of the $u$'s (by (6)), provided $\Delta u$ and $\Delta x$
contain no higher derivatives than the $\kappa$-th one occurring in
$f$.  Since $2 \kappa$-th derivatives in general occur in
$\psi$,\footnote{Provided $f$ is non-linear in the $\kappa$-th
derivatives.} therefore, we have the existence of $\rho$ first
integrals.  That there may be non-linear dependencies among these is
again shown by the above $f$.  To the linearly independent $\Delta u =
\epsilon_{1}$, $\Delta x = \epsilon_{2}$ there correspond the linearly
independent relations $\displaystyle{u^{\prime\prime} = \frac{d}{dx}
u^{\prime}}$; $\displaystyle{u^{\prime\prime} \cdot u^{\prime} =
\frac{1}{2} \frac{d}{dx} \left(u^{\prime} \right)^{2}}$; whereas
between the first integrals $u^{\prime} = \operatorname{const.}$, $u^{\prime
2} = \operatorname{const.}$ a non-linear dependency exists.  This relates to
the elementary case where $\Delta u$, $\Delta x$ contain no
derivatives of the $u$'s.\footnote{Otherwise we also have $u^{\prime
\lambda}=\operatorname{const.}$ for every $\lambda$, corresponding to
$$
u^{\prime \prime} \cdot (u^{\prime})^{\lambda -1} = \frac{1}{\lambda}
\frac{d}{dx} (u^{\prime})^{\lambda} .
$$
}

\section{Converse in Case of Infinite Group}

First let us show that the assumption of linearity of $\Delta x$ and
$\Delta u$ constitutes no restriction, a conclusion which follows,
even without the converse, from the fact that $\mathfrak{G}_{\infty
\rho}$ formally depends on $\rho$ and only $\rho$ arbitrary functions.
For it turns out that in the non-linear case, upon composition of the
transformations, whereby the terms of lowest order are added together,
the number of arbitrary functions would increase.  In fact, say, let
$$
y = A \left(x, u, \frac{\partial u}{\partial x}, \ldots ; p \right)
=x + \sum a(x, u, \ldots) p^{\nu} + 
b(x, u, \ldots) p^{\nu -1} \frac{\partial p}{\partial x} 
$$
$$
+ c \, p^{\nu-2} \left(\frac{\partial p}{\partial x} \right)^{2} + \ldots +
d \left(\frac{\partial p}{\partial x} \right)^{\nu} + \ldots \quad
\left(p^{\nu} = (p^{(1)})^{\nu_{1}} \ldots  
(p^{(\rho)})^{\nu_{\rho}} \right) ;
$$
and correspondingly $\displaystyle{v=B\left(x, u, \frac{\partial
u}{\partial x}, \ldots; p\right)}$; then by composition with
$\displaystyle{z=A\left(y, v, \frac{\partial v}{\partial y}, \ldots; q
\right)}$, for the terms of lowest order, we get
$$
z = x + \sum a (p^{\nu} + q^{\nu}) + 
b \left\{p^{\nu-1} \frac{\partial p}{\partial x} +   
q^{\nu-1} \frac{\partial q}{\partial x}  
\right\}
+ c \left\{p^{\nu-2} \left(\frac{\partial p}{\partial x} \right)^{2} 
+ q^{\nu-2} \left(\frac{\partial q}{\partial x} \right)^{2} 
\right\} + \ldots .
$$
%(Tavel p 196)
Here, if any coefficient different from $a$ and $b$ is different from
zero, in other words, if a term $\displaystyle{p^{\nu-\sigma} \left(
\frac{\partial p}{\partial x} \right)^{\sigma} + q^{\nu-\sigma} \left(
\frac{\partial q}{\partial x} \right)^{\sigma}}$ actually occurs for
$\sigma >1$ it cannot be written as a differential quotient of a
single function or power product of one; the number of arbitrary
functions, contrary to hypothesis, has thus increased.  If all
coefficients different from $a$ and $b$ vanish, then, according to the
values of the exponents $\nu_{1}$, $\ldots$, $\nu_{\rho}$, the second
term will become the differential quotient of the first (as always,
for example, for a $\mathfrak{G}_{\infty 1}$), so that linearity does
actually result; or else the number of arbitrary functions must again
increase.  The infinitesimal transformations, then, owing to the
linearity of the $p(x)$'s, satisfy a system of linear partial
differential equations; and since the group property is satisfied,
they constitute an ``infinite group of infinitesimal transformations''
accord to Lie's definition (Grundlagen, \S~10).

Now the converse is arrived at similarly to the case of the finite
group.  The fact that the dependencies (16) hold leads, through
multiplication by $p^{(\lambda)}$ and addition, by virtue of the
identity transformation (14), to $\sum \psi_{i} \overline{\delta}
u_{i} = \operatorname{Div} \Gamma$ and thence, as in Section 3 follows
the determination of $\Delta x$ and $\Delta u$ and the invariance of
$I$ with respect to these infinitesimal transformations, which do
actually depend linearly on $\rho$ arbitrary functions and their
derivatives up to the $\sigma$-th order.  The fact that these
infinitesimal transformations, if they
contain no derivatives $\displaystyle{\frac{\partial u}{\partial x}}$,
$\ldots$, certainly form a group, follows, as in Section 3, from the
fact that otherwise, by composition more arbitrary functions would
occur, whereas by assumption there are to be only $\rho$ dependencies
(16); hence they form an ``infinite group of infinitesimal
transformations.''  But such a one consists (Grundlagen, Theorem VII,
p. 391) of the most general infinitesimal transformations of a certain
``infinite group $\mathfrak{G}$ of finite transformations,'' in Lie's
sense, thereby defined.  Each such finite transformation is generated
from infinitesimal ones (Grundlagen, \S~7),\footnote{Hence it follows
in particular that the group $\mathfrak{G}$ generated from the
infinitesimal transformations $\Delta x$, $\Delta u$ of a
$\mathfrak{G}_{\infty \rho}$ reduces back to $\mathfrak{G}_{\infty
\rho}$.  For $\mathfrak{G}_{\infty \rho}$ contains no infinitesimal
transformations distinct from $\Delta x$, $\Delta u$ dependent on
arbitrary functions, and cannot contain any independent of them but
depending on parameters, as otherwise it would be a mixed group.  But
according to the above, the infinitesimal transformations determine
the finite ones.} and so arises through integration of the
simultaneous system
$$
\frac{d x_{i}}{dt} = \Delta x_{i} ; \quad
\frac{du_{i}}{dt} = \Delta u_{i} ; \quad
(x_{i}=y_{i}, \ u_{i} = v_{i} , \ \text{for} \ t = 0),
$$
where, however, it may be necessary further to assume the arbitrary
$p(x)$'s dependent on $t$.  Thus $\mathfrak{G}$ does actually depend
on $\rho$ arbitrary functions; if in particular it suffices to assume
$p(x)$ free from $t$, then this dependency becomes analytic in the
arbitrary function $q(x)=t \cdot p(x)$.\footnote{The question whether
perhaps this latter case always occurs was raised in a different
formulation by Lie (Grundlagen, \S~7 and \S~13 at end).} If
derivatives
%(Tavel p 197) 
$\displaystyle{\frac{\partial u}{\partial x}}$, $\ldots$, occurs, it
may be necessary also to adjoin infinitesimal transformation(s)
$\overline{\delta} u = 0$, $\operatorname{Div} (f \cdot \Delta x) = 0$
before drawing the same conclusions.  

In terms of an example of Lie's (Grundlagen, \S~7), let us add mention
of a fairly general case in which it is possible to break through to
explicit formulas, which at the same time show that the derivatives of
the arbitrary functions up to the $\sigma$-th order to occur; where,
in other words, the converse is complete.  I refer to such groups of
infinitesimal transformations of the $u$'s thereby ``induced''
corresponds; i.e., such transformations of the $u$'s for which $\Delta
u$, and consequently $u$, depend only on the arbitrary functions
occurring in $\Delta x$; assuming further that the derivatives
$\displaystyle{\frac{\partial u}{\partial x}}$, $\ldots$ do not occur
in $\Delta u$.  That is we have
$$
\Delta x_{i} = p^{(i)}(x); \quad \Delta u_{i} = \sum_{\lambda=1}^{n}
\left\{
a^{(\lambda)} (x,u) p^{(\lambda)} + 
b^{(\lambda)} \frac{\partial p^{(\lambda)}}{\partial x} + \ldots +
c^{(\lambda)} \frac{\partial^{\sigma} p^{(\lambda)}}{dx^{\sigma}}
\right\} .
$$
Since the infinitesimal transformation $\Delta x = p(x)$ generates
every transformation $\Delta x = y + g(y)$ with arbitrary $g(y)$, we
can in particular determine $p(x)$ to depend on $t$ in such a matter
as to generate the single-member group
\begin{equation}
x_{i} = y_{i} + t \cdot g_{i}(y) ,
\end{equation}
which goes over into the identity for $t=0$ and into the required
$x=y+g(y)$ for $t=1$.  For by differentiation of (18), it follows that
\begin{equation}
\frac{d x_{i}}{d t} = g_{i}(y) = p^{(i)}(x,t), 
\end{equation}
where $p(x,t)$ is determined from $g(y)$ by inversion of (18); and
conversely, (18) is generated from (19) by virtue of the auxiliary
condition $x_{i}=y_{i}$ for $t=0$, by which the integral is uniquely
determined.  By means of (18), the $x$'s can be replaced in $\Delta u$
by the ``constants of integration'' $y$ and by $t$; the $g(y)$'s
occurring just up to the $\sigma$-th derivative, the
$\displaystyle{\frac{\partial y}{\partial x}}$'s being expressed in
terms of $\displaystyle{\frac{\partial x}{\partial y}}$ in
$\displaystyle{\frac{\partial p}{\partial x} = \sum \frac{\partial
g}{\partial y_{\kappa}} \frac{\partial y_{\kappa}}{\partial x}}$, and
$\displaystyle{\frac{\partial^{\sigma} p}{\partial x^{\sigma}}}$ being
in general replaced by its value in $\displaystyle{\frac{\partial
g}{\partial y}}$, $\ldots$, $\displaystyle{\frac{\partial x}{\partial
y}}$, $\ldots$, $\displaystyle{\frac{\partial^{\sigma} x}{\partial
y^{\sigma}}}$.  For the determination of the $u$'s we thus obtain the
system of equations
$$
\frac{d u_{i}}{dt} = F_{i} \left( g(y), \frac{\partial g}{\partial y} ,
\ldots \frac{\partial^{\sigma} g}{\partial y^{\sigma}}, u, t
\right) \quad
(u_{i} = v_{i} \ \text{for} \ t = 0)
$$
in which only $t$ and $u$ are variables, while the $g(y)$, $\ldots$
pertain to the field of coefficients, so that integration yields
%(Tavel p 198)
$$
u_{i} = v_{i} + B_{i} \left(v, g(y), \frac{\partial g}{\partial y},
\ldots \frac{\partial^{\sigma} g}{\partial y^{\sigma}} , t
\right)_{t=1} ,
$$
or transformations depending on exactly $\sigma$ derivatives of the
arbitrary functions.  The identity is contained in this, by (18), for
$g(y)=0$; and the group property follows from the fact that the method
specified affords every transformation $x = y + g(y)$, whereby the
induced transformation of the $u$'s is uniquely determined, and the
group $\mathfrak{G}$ accordingly exhausted.

From the converse it follows incidentally that it constitutes no
restriction to assume the arbitrary functions to be dependent only on
the $x$'s, not on the $u$, $\displaystyle{\frac{\partial u}{\partial
x}}$, $\ldots$.  For in the latter event, the identity transformation
(14), and hence also (15), would involve not only the
$p^{(\lambda)}$'s but also $\displaystyle{\frac{\partial
p^{(\lambda)}}{\partial u}}$, $\displaystyle{\frac{\partial
p^{(\lambda)}}{\partial \frac{\partial u}{\partial x}, \ldots}}$.  Now
if we assume the $p^{(\lambda)}$'s to be successively of the zeroth,
first, $\ldots$ degree in $u$, $\displaystyle{\frac{\partial
u}{\partial x}}$, $\ldots$, with arbitrary functions of $x$ as
coefficients, then we again obtain dependencies (16), only in greater
number; which, however, according to the above converse, through
conjunction with arbitrary functions dependent on $x$ only, reduce to
the previous case.  In the same way it is shown that mixed groups
correspond to simultaneous occurrence of dependencies and of
divergence relationships independent of them.\footnote{As in Section
3, it here again follows from the converse that besides $I$, every
integral $I^{\ast}$ different from it by an integral over a divergence
likewise admits of an infinite group, with the same $\overline{\delta}
u$'s, though $\Delta x$ and $\Delta u$ will in general involve
derivatives of the $u$'s.  Such an integral $I^{\ast}$ was introduced
by Einstein in the general theory of relativity to obtain a simpler
version of laws of conservation of energy; I specify the
infinitesimal transformations that this $I^{\ast}$ admits of, adhering
precisely in nomenclature to Klein's second Note.  The integral $I=
\int \ldots \int K \, d \omega = \int \ldots \int \mathfrak{K} \, d S
$ admits of the group of \emph{all} transformations of the $\omega$'s
and those induced thereby for the $g_{\mu \nu}$'s; to this correspond
the dependencies (Klein's (30))
$$ \sum \mathfrak{K}_{\mu \nu} g_{\tau}^{\mu \nu} + 
2 \sum \frac{\partial g^{\mu \nu} 
\mathfrak{K}_{\mu \tau}}{\partial \omega^{\sigma}} = 0. 
$$
Now $I^{\ast} = \int \ldots \int \mathfrak{K}^{\ast} \, d S$, where
$\mathfrak{K}^{\ast} = \mathfrak{K} + \operatorname{Div}$, and
consequently $\mathfrak{K}_{\mu \nu}^{\ast} = \mathfrak{K}_{\mu \nu}$,
where $\mathfrak{K}_{\mu \nu}^{\ast}$, $\mathfrak{K}_{\mu \nu}$ stand
in each instance for the Lagrange expressions.  The dependencies
specified are therefore such for $\mathfrak{K}_{\mu \nu}^{\ast}$ also;
and after multiplication by $p^{\tau}$ and addition, we obtain,
applying the transformations of product differentiation in reverse,
$$
\sum \mathfrak{K}_{\mu \nu} p^{\mu \nu} + 2 \, \operatorname{Div}
\left( \sum g^{\mu \sigma} \mathfrak{K}_{\mu \tau} p^{\tau}
\right) = 0 ;
$$
$$
\delta \mathfrak{K}^{\ast} + \operatorname{Div} \sum
\left(
2 g^{\mu \sigma} \mathfrak{K}_{\mu \tau} p^{\tau} -
\frac{\partial \mathfrak{K}^{\ast}}{\partial g_{\sigma}^{\mu \nu}}
p^{\mu \nu} \right) = 0 .
$$
Comparing this with Lie's differential equation 
$\delta \mathfrak{K}^{\ast} + \operatorname{Div} (\mathfrak{K}^{\ast} 
\Delta \omega) = 0$,
$$
\Delta \omega^{\sigma} = \frac{1}{\mathfrak{K}^{\ast}} \cdot
\sum \left(
2 g^{\mu \sigma} \mathfrak{K}_{\mu \tau} p^{\tau} -
\frac{\partial \mathfrak{K}^{\ast}}{\partial g_{\sigma}^{\mu \nu}}
p^{\mu \nu}
\right); \quad
\Delta g^{\mu \nu} = p^{\mu \nu} + \sum g_{\sigma}^{\mu \nu} \Delta
\omega^{\sigma}
$$
follow as infinitesimal transformations of which $I^{\ast}$ admits.
These infinitesimal transformations, then depend on the first and
second derivatives of the $g^{\mu \nu}$'s, and contain the arbitrary
$p$'s as far as the first derivative.}

\section{Invariance of the Several Constituents of the Relations}

If we specialize the group $\mathfrak{G}$ to be the simplest case
usually considered by allowing no derivatives of the $u$'s in the
transformations, and in that the transformed independent variables
depend only on the $x$'s not on the $u$'s, we can infer invariance of
the several constituents in formulas.  To begin with, by known
arguments, we get invariance of $\int \ldots \int (\sum \psi \delta
u_{i}) \, dx$; relative invariance, that is of $\sum \psi_{i} \delta
u_{i}$,\footnote{That is, $\sum \psi_{i} \delta u_{i}$ takes on a
factor upon transformation, and this always used to be termed relative
invariance in the algebraic theory of invariance.} meaning by $\delta$
any variation.  For we have in the first place
$$
\delta I = \int \ldots \int \delta 
f \left(x, u, \frac{\partial u}{\partial x}, \ldots \right) dx
= \int \ldots \int \delta 
f \left(y, v, \frac{\partial v}{\partial y}, \ldots \right) dy ,
$$
and in the second place, for $\delta u$, $\displaystyle{\delta
\frac{\partial u}{\partial x}}$, $\ldots$ vanishing at the boundary,
according to which $\delta v$, $\displaystyle{\delta \frac{\partial
v}{\partial y}}$, $\ldots$ vanishing at the boundary also owing to the
linear homogeneous transformation of the $\delta u$,
$\displaystyle{\delta \frac{\partial u}{\partial x}}$, $\ldots$,
$$
\int \ldots \int \delta f \left(x, u, 
\frac{\partial u}{\partial x}, \ldots \right) dx  =
\int \ldots \int  \left( \sum \psi_{i}( u, \ldots) \delta u_{i} 
\right) d x;
$$
%Tavel p 199
$$
\int \ldots \int \delta f \left(y, v, 
\frac{\partial v}{\partial y}, \ldots \right) dy  =
\int \ldots \int  \left( \sum \psi_{i}( v, \ldots) \delta v_{i} 
\right) d y,
$$
and consequently, for $\delta u$, $\displaystyle{\delta \frac{\partial
u}{\partial x}}$, $\ldots$ vanishing at the boundary,
$$
\int \ldots \int \left( \sum \psi_{i} (u, \ldots) \delta u_{i} \right)
dx =
\int \ldots \int \left( \sum \psi_{i} (v, \ldots) \delta v_{i} \right)
dy 
$$
$$
=
\int \ldots \int \left( \sum \psi_{i} (v, \ldots) \delta v_{i} \right)
\left| \frac{\partial y_{i}}{\partial x_{\kappa}} \right| dx .
$$
If in the third integral $y$, $v$, $\delta v$ are expressed in terms
of $x$, $u$, $\delta u$, and the third is equated to the first, we
thus have a relationship
$$
\int \ldots \int \left(\sum \chi_{i} (u, \ldots) \delta u_{i} \right) 
dx= 0
$$
for $\delta u$ vanishing at the boundary but otherwise arbitrary, and
thence follows, familiarly, the vanishing of the integrand for any
$\delta u$ whatever; the relation
$$
\sum \psi_{i} ( u, \ldots) \delta u_{i} = 
\left|\frac{\partial y_{i}}{\partial x_{\kappa}} \right|
\left( \sum \psi_{i} (v, \ldots) \delta v_{i} \right) ,
$$
identical in $\delta u$, therefore holds, asserting the relative
invariance of $\sum \psi_{i} \delta u_{i}$ and consequently the
invariance of $\int \ldots \int \left(\sum \psi_{i} \delta u_{i}
\right) dx$.\footnote{These conclusions fail if $y$ depends also on
the $u$'s, since in that case $\displaystyle{\delta f \left(y,v,
\frac{\partial v}{\partial y}, \ldots \right)}$ also contains terms
$\displaystyle{\sum \frac{\partial f}{\partial y} \delta y}$, so that
the divergence transformation does not lead to the Lagrange
expressions; and similarly if derivatives of the $u$'s are admitted;
for in that case the $\delta v$'s become linear combinations of
$\delta u$, $\displaystyle{\delta \frac{\partial u}{\partial x}}$,
$\ldots$, and so lead only after another divergence transformation to
an identity $\int \ldots \int \left(\sum \chi_{i}(u, \ldots) \delta u
\right) dx=0$, so that again the Lagrange expressions do not appear on
the right.

The question whether it is possible to argue from the invariance of
$\int \ldots \int \left(\sum \psi_{i} \delta u_{i} \right) d x$ back
to the subsistence of divergence relationships is synonymous,
according to the converse, with the question whether one can thence
infer the invariance of $I$ with respect to a group leading not
necessarily to the same $\Delta u$, $\Delta x$, but to the same
$\overline{\delta} u$'s.  In the special case of the single integral
and only first derivatives in $f$, it is possible for the finite group
to argue from the invariance of the Lagrange expressions to the
existence of first integrals (c.f. e.g., Engel, G\"{o}tt. Nachr. 1916,
p. 270).}

To apply this to the divergence relationships and dependencies
derived, we must first demonstrate that the $\overline{\delta} u$
derived from the $\Delta u$, $\Delta x$'s does in fact satisfy the
laws of transformation for the variation $\delta u$, provided only
that the parameters, or arbitrary functions, in $\overline{\delta} v$
are so determined as corresponds to the similar group of infinitesimal
transformations in $y$, $v$; if $\mathfrak{T}_{q}$ designates the
transformation that carries $x$, $u$ over into $y$, $v$, and
$\mathfrak{T}_{p}$ and infinitesimal one in $x$, $u$, then the one
similar thereto in $y$, $v$ is given by $\mathfrak{T}_{r} =
\mathfrak{T}_{q} \mathfrak{T}_{p} \mathfrak{T}_{q}^{-1}$, where the
parameters, or arbitrary functions $r$, are thus determined from $p$
and $q$.  In formulas, this is expressed as follows:
$$
\mathfrak{T}_{p}: \xi = x + \Delta x (x,p); \quad 
u^{\ast} = u + \Delta u(x,u,p);
$$
$$
\mathfrak{T}_{q}: y = A(x,q); \quad
v = B(x, u, q);
$$
$$
\mathfrak{T}_{q} \mathfrak{T}_{p}: \eta = A(x + \Delta x(x,p), \, q); \quad
v^{\ast} = B(x + \Delta x(p), u + \Delta u(p), \, q).
$$
But this generates $\mathfrak{T}_{r} = \mathfrak{T}_{q}
\mathfrak{T}_{p} \mathfrak{T}_{q}^{-1}$, or
$$
\eta = y + \Delta y(r); \quad v^{\ast} = v + \Delta v(r) ,
$$
if by the inverse $\mathfrak{T}_{q}$ we regard the $x$'s as functions
of the $y$'s and consider only the infinitesimal terms; so we have the
identity
%Tavel p 200%
$$
\eta = y + \Delta y(r) = y + \sum \frac{\partial A(x,q)}{\partial x}
\Delta x(p);
$$
\begin{equation}
v^{\ast} = v + \Delta v(r) = v + \sum \frac{\partial B(x,u,q)}{\partial x}
\Delta x(p) + \sum \frac{\partial B(x,u,q)}{\partial u} \Delta u(p) .
\end{equation}
Replacing $\xi=x+ \Delta x$ by $\xi - \Delta \xi$ in this so that
$\xi$ goes back into $x$, and $\Delta x$ vanishes, by the first
equation (20) $\eta$ too will go back over into $y=\eta - \Delta
\eta$; if by this substitution $\Delta u(p)$ goes over into
$\overline{\delta} u(p)$, then $\Delta v(r)$ will go over into
$\overline{\delta} v(r)$ as well, and the second formula (20) gives
$$
v + \overline{\delta} v(y,v, \ldots r) = v + \sum 
\frac{\partial B(x,u,q)}{\partial u} \overline{\delta} u (p),
$$
$$
\overline{\delta} v(y,v,\ldots r) = \sum 
\frac{\partial B}{\partial u_{\kappa}} \overline{\delta} 
u_{\kappa} (x, u, p),
$$
so that the transformation formulas for variations are actually
satisfied provided $\overline{\delta} v$ is assumed to depend only on
the parameters or arbitrary functions $r$.\footnote{It turns out again
that $y$ must be taken independent of $u$ in order for the conclusions
to hold.  As an example, consider the $\delta g^{\mu \nu}$ and $\delta
q_{\rho}$ given by Klein, which satisfy the transformations for
variations provided the $p$'s are subjected to a vector
transformation.}

So in particular, the relative invariance of $\sum \psi_{i}
\overline{\delta} u_{i}$ follows: hence also, by (12), since the
divergence relationships are satisfied in $y$, $v$ as well, the
relative invariance of $\operatorname{Div} B$; and further, by (14)
and (13), the relative invariance of $\operatorname{Div} \Gamma$ and
of the left-hand sides of the dependencies as conjoined with the
$p^{(\lambda)}$'s, where the arbitrary $p(x)$'s (or the parameters)
are to be replaced by the $r$'s everywhere in the transformed
formulas.  This leads as well to the relative invariance of
$\operatorname{Div} (B - \Gamma)$, or of a divergence of a not
identically vanishing system of functions $B-\Gamma$ whose divergence
vanishes identically.

From the relative invariance of $\operatorname{Div} B$, we can draw
additional inference of invariance of the first integral in the
one-dimensional case and for finite group.  The parametric
transformation corresponding to the infinitesimal transformation
becomes, by (20), linear and homogeneous, and owing to the
invertibility of all transformations, the $\epsilon$'s also will be
linear and homogeneous in the transformed parameters
$\epsilon^{\ast}$.  This invertibility is certainly preserved if we
put $\psi=0$, since no derivatives of the $u$'s occur in (20).
Through equating the coefficients of the $\epsilon^{\ast}$'s in
$$
\operatorname{Div} B(x,u, \ldots \epsilon) = \frac{dy}{dx} \cdot
\operatorname{Div} B(y, v, \ldots \epsilon^{\ast})
$$
%Tavel p 201
the $\displaystyle{\frac{d}{dy} B^{(\lambda)} (y, v, \ldots)}$'s
therefore also become linear homogeneous functions of
$\displaystyle{\frac{d}{dx} B^{(\lambda)} (x, u, \ldots)}$'s so that
$\displaystyle{\frac{d}{dx} B^{(\lambda)} (x, u, \ldots)} = 0$ or
$B^{(\lambda)}(x,u) = \operatorname{const.}$ duly entails
$\displaystyle{\frac{d}{dy} B^{(\lambda)} (y, v, \ldots)} = 0$ or
$B^{(\lambda)}(y,v) = \operatorname{const.}$ as well.  So the $\rho$ first
integrals corresponding to a $\mathfrak{G}_{\rho}$ in each instance
admit of the group, with result that the further integration is
simplified.  The simplest example of this is that $f$ is free of $x$
or of a $u$, which corresponds to the infinitesimal transformation
$\Delta x=\epsilon$, $\Delta u = 0$, or $\Delta x = 0$, $\Delta u =
\epsilon$.  We shall have $\displaystyle{\overline{\delta} u = -
\epsilon \frac{du}{dx}}$ or $\epsilon$ respectively, and since $B$ is
derived from $f$ and $\overline{\delta} u$ by differentiation and
rational combination, it is free accordingly of $x$ or $u$
respectively, and admits of the corresponding groups.\footnote{In the
cases where mere invariance of $\int \left(\sum \psi_{i} \delta u_{i}
\right) dx$ entails the existence of first integrals, these do not
admit of the entire group $\mathfrak{G}_{\rho}$; for example, $\int
(u^{\prime \prime} \delta u) dx$ admits of the infinitesimal
transformation $\Delta x = \epsilon_{2}$, $\Delta u = \epsilon_{1} + x
\epsilon_{3}$; whereas the first integral $u - u^{\prime} x =
\operatorname{const.}$, corresponding to $\Delta x=0$, $\Delta u = x
\epsilon_{3}$, does not admit of the other two infinitesimal
transformations, since it explicitly contains both $u$ and $x$.  To
this first integral, there happen to correspond infinitesimal
transformations for $f$ that contain derivatives.  So we see that
invariance $\int \ldots \int \left(\sum \psi_{i} \delta u_{i} \right)
dx$ is at all events a weaker condition than invariance of $I$, and
this should be noted as to a question raised in a previous remark.}

\section{A Hilbertian Assertion}

From the foregoing, finally, we also obtain the proof of a Hilbertian
assertion about the connection of the failure of laws of conservation
of energy proper with ``general relativity'' (Klein's first Note,
G\"{o}ttinger Nachr. 1917, Reply 1st paragraph), and that in a
generalized group theory version.

Let the integral $I$ admit of a $\mathfrak{G}_{\infty \rho}$, and let
$\mathfrak{G}_{\rho}$ be any finite group generated by specializing
the arbitrary functions, and hence a subgroup of $\mathfrak{G}_{\infty
\rho}$.  Then to the infinite group $\mathfrak{G}_{\infty \rho}$ there
correspond dependencies (16), and to the finite one
$\mathfrak{G}_{\sigma}$, divergence relationships (13); and conversely
from the subsistence of any divergence relationships, the invariance
of $I$ follows, with respect to some finite group which will be
identical with $\mathfrak{G}_{\sigma}$ if and only if the
$\overline{\delta} u$'s are linear combinations of those obtained from
$\mathfrak{G}_{\sigma}$.  Thus the invariance with respect to
$\mathfrak{G}_{\sigma}$ cannot lead to any divergence relationships
different from (13).  But since the subsistence of (16) entails the
invariance of $I$ with respect to the infinitesimal transformations,
$\Delta u$, $\Delta x$ of $\mathfrak{G}_{\infty \rho}$ for \emph{any}
$p(x)$, it entails in particular nothing less than invariance with
respect to the infinitesimal transformations of
$\mathfrak{G_{\sigma}}$ arising therefrom by specialization and
consequently with respect to $\mathfrak{G}_{\sigma}$.  Thus the
divergence relationships $\sum \psi_{i} \overline{\delta}
u_{i}^{(\lambda)} = \operatorname{Div} B^{(\lambda)}$ must be
consequences of the dependencies (16), which latter may alternatively
be written $\sum \psi_{i} a_{i}^{(\lambda)} = \operatorname{Div}
\chi^{(\lambda)}$ where the $\chi^{(\lambda)}$'s are linear
combinations of the Lagrange expressions and their derivatives.  Since
the $\psi$'s occur linearly in both (13) and (16), the divergence
relations must thus in particular be \emph{linear} combinations of the
dependencies (16);
%Tavel p 202
Accordingly, $\displaystyle{\operatorname{Div} B^{(\lambda)} =
\operatorname{Div} \left(\sum \alpha \cdot \chi^{(\kappa)} \right) }$;
and the $B^{(\lambda)}$'s themselves are thus linearly composed of the
$\chi$'s, i.e., the Lagrange expressions and their derivatives, and of
functions whose divergence vanishes identically, say like the
$B-\Gamma$'s encountered at the close of Section 2, for which
$\operatorname{Div} (B- \Gamma) = 0$, and where the divergence at the
same time has the invariant property.  I shall refer to divergence
relationships in which the $B^{(\lambda)}$'s can be composed from the
Lagrange expressions and their derivatives in the specified manner as
``improper,'' and to all other as ``proper.''

If conversely the divergence relations are linear combinations of the
dependencies (16), and so ``improper,'' invariance with respect to
$\mathfrak{G}_{\sigma}$ follows from that with respect to
$\mathfrak{G}_{\infty \rho}$; $\mathfrak{G}_{\sigma}$ becomes a
subgroup of $\mathfrak{G}_{\infty \rho}$.  The divergence
relationships corresponding to an infinite group
$\mathfrak{G}_{\sigma}$ will thus be improper if and only if
$\mathfrak{G}_{\sigma}$ is a subgroup of an infinite group invariant
with respect to $I$.

By specialization of the groups, this yields the original Hilbertian
assertion.  Let ``displacement group'' be understood to mean the
finite
$$
y_{i} = x_{i} + \epsilon_{i} ; \quad 
v_{i}(y) = u_{i}(x) ;
$$
that is
$$
\Delta x_{i} = \epsilon_{i} , \quad
\Delta u_{i} = 0 , \quad
\overline{\delta} u_{i} = - \sum_{\lambda} 
\frac{\partial u_{i}}{\partial x _{\lambda}} \epsilon_{\lambda} .
$$
Invariance with respect to the displacement group asserts, as we know,
that in \newline
$\displaystyle{I=\int \ldots \int f \left(x, u, \frac{\partial
u}{\partial x}, \ldots \right) dx}$, the $x$'s do not occur explicitly
in $f$.  The associated $n$ divergence relationships
$$
\sum \psi_{i} \frac{\partial u_{i}}{\partial x_{\lambda}} =
\operatorname{Div} B^{(\lambda)} \quad (\lambda = 1, 2, \ldots n)
$$
will be referred to as ``energy relationships,'' since the laws of
conservation'' $\operatorname{Div} B^{(\lambda)} = 0$ corresponding to
the variation problem answer to ``laws of conservation of energy,''
and the $B^{(\lambda)}$'s to the ``energy components.''  So then we
have: If $I$ admits of the displacement group, then the energy
relationships become improper if and only if $I$ is invariant with
respect to an infinite group containing the displacement group as
subgroup.\footnote{The laws of conservation of energy of classical
mechanics as well as those of the old ``theory of relativity'' (where
$\sum d x^{2}$ goes over into itself) are ``proper,'' since no
infinite groups occur.}

An example of such infinite groups is presented by the group of
\emph{all} transformations of the $x$'s and such of the induced
transformations of the $u(x)$'s in which only \emph{derivatives} of
the arbitrary functions $p(x)$ occur; the displacement group is
generated by the specialization $p^{(i)}(x) = \epsilon_{i}$; but it
must remain undecided whether this --- and the groups generated by
change of $I$ by a boundary integral --- suffices to give the most
general of these groups.  Induced transformations of the specified
kind arise, say, when the $u$'s are subjected to the coefficient
transformations of a ``total differential form,'' i.e., a form $\sum a
\, d^{\lambda} x_{i} + \sum b \, d^{\lambda -1} x_{i} d x_{\kappa} +
\ldots$ containing higher differentials besides the $dx$'s; more
special induced transformations, in which the $p(x)$'s occur in first
derivative only, are furnished by the coefficient transformations of
ordinary differential forms $\sum c \, dx_{i_{1}} \ldots d
x_{i_{\lambda}}$, and only these have ordinarily been considered.

Another group of the specified kind --- one which, owing to the
occurrence of the logarithmic term, cannot be coefficient
transformation --- is, say, the following:
$$
y = x + p(x); \quad v_{i} = u_{i} + \ln (1 + p^{\prime}(x)) =
u_{i} + \ln \frac{dy}{dx} ;
$$
$$
\Delta x = p(x); \quad \Delta u_{i} = p^{\prime}(x) ;\footnote{From 
these infinitesimal transformations, the finite ones are
calculated backwards by the method given in Section 4 at end.} 
\quad 
\overline{\delta} u_{i} = p^{\prime}(x) - u_{i}^{\prime} p(x).
$$
The dependencies (16) here become
$$
\sum_{i} 
\left( \psi_{i} u_{i}^{\prime} + \frac{d \psi_{i}}{dx} \right) = 0,
$$
and the improper energy relationships
$$
\sum \left(\psi_{i} u_{i}^{\prime} + 
\frac{d (\psi_{i} + \operatorname{const.})}{dx} \right) = 0.
$$
A simple invariant integral of the group is
$$
I = \int \frac{e^{-2u_{1}}}{u_{1}^{\prime} - u_{2}^{\prime}} d x.
$$
The most general $I$ is determined by integration of Lie's differential
equation (11)
$$
\overline{\delta} f + \frac{d}{dx} (f \cdot \Delta x) = 0 ,
$$
which by substitution of their values for $\Delta x$ and
$\overline{\delta} u$, provided $f$ is assumed to depend on only first
derivatives of the $u$'s, goes over into
%Tavel p 204
$$
\frac{\partial f}{\partial x} p(x) + \left\{
\sum \frac{\partial f}{\partial u_{i}} - 
\frac{\partial f}{\partial u_{i}^{\prime}} u_{i}^{\prime} +
f \right\} p^{\prime}(x) + 
\left\{ \sum \frac{\partial f}{\partial u_{i}^{\prime\prime}} 
\right\} p^{\prime\prime}(x) = 0
$$
(identically in $p(x)$, $p^{\prime}(x)$, $p^{\prime\prime}(x))$.  This
system of equations has solutions for as few as two functions $u(x)$
actually containing the derivatives, namely
$$
f = (u_{1}^{\prime} - u_{2}^{\prime}) \, \varPhi \! \left(u_{1} - u_{2},
\frac{e^{-u_{1}}}{u_{1}^{\prime} - u_{2}^{\prime}} \right) ,
$$
where $\varPhi$ stands for an arbitrary function of the specified
arguments.

Hilbert enunciates his assertion to the effect that the failure of
proper laws of conservation of energy is a characteristic feature of
the ``general theory of relativity.''  In order for this assertion to
hold good literally, therefore, the term ``general relativity'' should
be taken in a broader sense than usual, and extended also to the
forgoing groups depending on $n$ arbitrary functions.\footnote{This
again confirms the correctness of a comment of Klein's that the term
``relativity'' current in physics is replaceable by ``invariance
relative to a group.''  (``\"{U}ber die geometrishen Grundlagen der
Lorentzgruppe,'' Jhrber. d. Deutsch. Math. Vereinig. \textbf{19}
(1910), p. 287, reprinted in the Phys. Zeitschrift.)}

\end{document}